\documentclass[hidelinks,letter, 10 pt, conference,doublecolomn]{ieeeconf}

\IEEEoverridecommandlockouts    
\usepackage{balance}
\usepackage{mathtools}
\usepackage{siunitx}
\title{\LARGE \bf
Input-Output Extension of Underactuated Nonlinear Systems}

\usepackage{mathrsfs} \usepackage{bm}
\usepackage{subfiles}
\usepackage{graphicx}
\usepackage{epstopdf}
\usepackage{float}

\usepackage{bm}
\usepackage{amsmath}
\usepackage{amssymb}

\newcommand{\vect}[1]{\mathbf{#1}}
\newcommand{\mat}[1]{\mathbf{#1}}

\newcommand{\diffs}[3]{\frac{\partial^2 #1}{
\ifx#2#3 
\partial #2^2
\else
\partial #2 \partial #3
\fi
}}

\newcommand{\alphav}{\boldsymbol{\alpha}}

\newcommand{\betav}{\bm{\beta}}

\newcommand{\av}{\vect{a}}
\newcommand{\bv}{\vect{b}}
\newcommand{\cv}{\vect{c}}

\newcommand{\dv}{\vect{d}}
\newcommand{\ev}{\vect{e}}

\newcommand{\fv}{\vect{f}}
\newcommand{\gv}{\vect{g}}

\newcommand{\hv}{\vect{h}}

\newcommand{\lv}{\vect{l}}

\newcommand{\nv}{\vect{n}}

\newcommand{\pv}{\vect{p}}
\newcommand{\dpv}{\dot{\vect{p}}}

\newcommand{\xiv}{\bm{\xi}}

\newcommand{\ddpv}{\ddot{\vect{p}}}

\newcommand{\rv}{{\vect{r}}}

\newcommand{\uv}{\vect{u}}
\newcommand{\vv}{\vect{v}}

\newcommand{\wv}{\vect{w}}

\newcommand{\xv}{\vect{x}}
\newcommand{\dxv}{\dot{\vect{x}}}

\newcommand{\yv}{\vect{y}}

\newcommand{\zv}{\vect{z}}

\newcommand{\etav}{\bm{\eta}}
\newcommand{\varphiv}{\bm{\varphi}}

\newcommand{\SO}{\mathbf{SO}(3)}

\newcommand{\rhov}{\bm{\rho}}

\newcommand{\Omegav}{\bm{\Omega}}

\newcommand{\Phim}{\bm{\Phi}}

\newcommand{\tauv}{\bm{\tau}}

\newcommand{\Gammam}{\bm{\Gamma}}

\newcommand{\Am}{\mat{A}}

\newcommand{\Gm}{\mat{G}}

\newcommand{\Jm}{\mat{J}}
\newcommand{\Lm}{\mat{L}}

\newcommand{\Km}{\mat{K}}

\newcommand{\Rm}{\mat{R}}

\newcommand{\Tm}{\mat{T}}

\usepackage{svg}
\usepackage{hyperref}
\usepackage{academicons}
\usepackage{xcolor}
\newcommand{\update}[1]{{\color{magenta}}}

\usepackage{amsthm}
\theoremstyle{plain}
\newtheorem{defn}{Definition}

\newtheorem{thm}{Theorem}

\newtheorem{problem}{Problem}

\usepackage{comment}
\usepackage[font=small]{caption}
\usepackage{subcaption}
\usepackage{calc}
\newtheorem{rem}{\textbf{Remark}}[section]
\usepackage{mathtools} 

\usepackage{mathptmx}
\usepackage{orcidlink}

\author{Mirko Mizzoni$^1$\orcidlink{0009-0006-2165-3475},  
Amr Afifi$^1$\orcidlink{0000-0002-2267-575X}, and Antonio Franchi$^{1,2}$\orcidlink{0000-0002-5670-1282}
\thanks{$^1$Robotics and Mechatronics group, Faculty of Electrical Engineering,  Mathematics, and Computer Science (EEMCS), University of Twente, 7500 AE Enschede, The Netherlands. {\footnotesize \tt m.mizzoni@utwente.nl}, {\footnotesize \tt a.n.m.g.afifi@utwente.nl}, {\footnotesize \tt a.franchi@utwente.nl}}
\thanks{$^2$Department of Computer, Control and Management Engineering, Sapienza University of Rome, 00185 Rome, Italy, {\footnotesize \tt antonio.franchi@uniroma1.it}}\thanks{This work was partially funded by the Horizon Europe research agreement no. 101120732 (AUTOASSESS).}}

\newif\ifappendix
\appendixfalse 

\begin{document}

\maketitle
\thispagestyle{empty}
\pagestyle{empty}

\begin{abstract}
This letter proposes a method to integrate auxiliary actuators that enhance the task-space capabilities of commercial underactuated systems leaving internal  certified low-level controller untouched.
The additional actuators are combined with a feedback-linearizing outer loop controller, enabling full-pose tracking. We provide the conditions under which legacy high-level commands  and new actuator inputs can be cohesively coordinated to achieve decoupled control of all degrees of freedom.
A comparative study with a standard quadrotor—originally not designed for physical interaction—demonstrates that the proposed modified platform remains stable under contact, while the baseline system diverges.
Additionally, simulation results under parameter uncertainty illustrate the approach’s robustness.
\end{abstract}

\section{Introduction}
Under-actuated (UA) mechanical systems are  characterized by  fewer actuators than the degrees of Freedom (DoF)~\cite{doi:10.1177/1729881419862164,survey}.   These systems are prevalent in robotics and aerospace engineering. A widely studied example is the quadrotor platform, which has gained prominence in applications like aerial surveillance, delivery systems, and environmental monitoring due to its cost efficiency, mechanical simplicity, and agility in unstructured environments.

 Robotic tasks such as manipulation, require full six-degree-of-freedom (6D) control, pushing the limits of what underactuated systems can achieve with their original actuation and control architectures.
For example, in aerial physical interaction (APhI) tasks—where an aerial robotic system must interact physically with the environment—underactuated vehicles like quadrotors face inherent limitations ~\cite{2019h-RylMusPieCatAntCacBicFra}. It has been shown~\cite{NGUYEN2015289} that when a quadrotor is required to control the tip position of a rigidly attached tool, the system may exhibit an unstable internal dynamics.

\begin{figure}[t]
    \centering
\includegraphics[width=0.51\textwidth,  trim=0cm 0cm 0cm 0cm, clip]{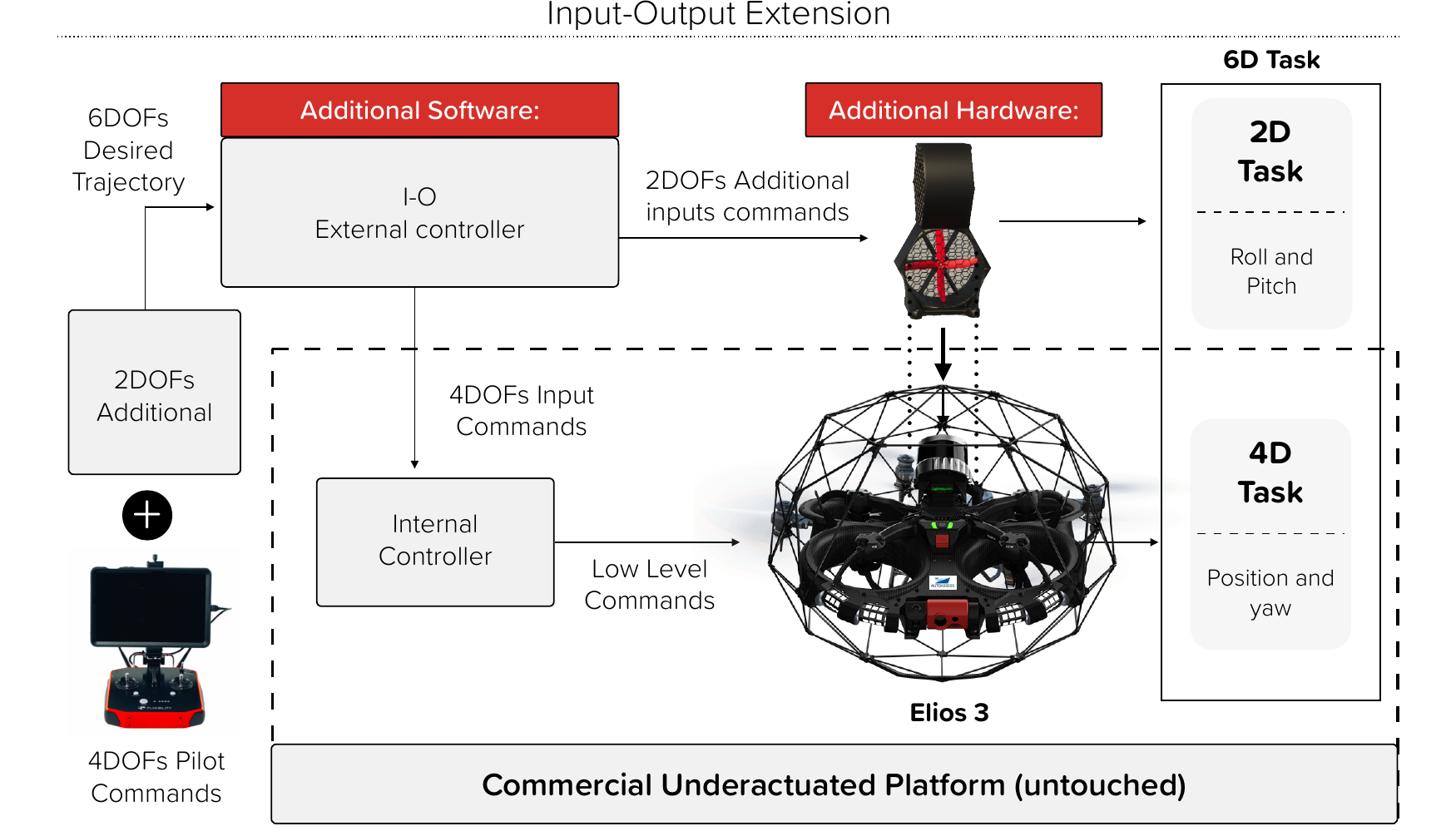}
     \caption{A quadrotor developed for industrial inspection in confined spaces  with a built-in controller, the "Flyability Elios 3" designed to follow high-level motion commands—is retrofitted with two additional rotors. These new actuators are not accounted for by the internal control law. We show that, under suitable conditions, feedback linearization can reinterpret the original high-level commands and integrate the additional inputs to achieve full 6D pose tracking. This is done without modifying the internal controller, offering practical and cost-effective advantages.} 
    \label{fig:quad}
\end{figure}

The robotics and control communities have responded to these challenges in several ways. One line of research focuses on the development of fully actuated platforms, such as hexarotors with tilted rotors~
\cite{2016j-RylBicFra,2016-GirSanGhe,9815278},
which can achieve arbitrary force and torque generation in all directions. However, these solutions often involve significant hardware redesign and the loss of commercial advantages such as certification, reliability, and ease of use.

Despite significant progress in underactuated control, including energy-based methods, backstepping, and hybrid control strategies~\cite{10854614,9416831},~\cite{2010-LeeLeoMcc,cite-fdl}, many existing approaches assume full access to low-level actuation or system models—assumptions that do not hold in commercial, closed-source platforms. Moreover, much of the existing literature addresses stabilization and trajectory tracking, rather than expanding the system’s functional output space.

This letter addresses a key gap in the state of the art: how to endow general commercial underactuated systems with enhanced task-space capabilities by adding auxiliary actuators, while preserving the integrity and certification of their internal controllers. We focus on the novel  problem of \emph{input-output extension}, where additional actuators are integrated at the system level and coordinated through a feedback-linearizing outer loop. This approach allows the original high-level commands—e.g., velocity or yaw references in a commercial quadrotor —to be reinterpreted, enabling the platform to control previously inaccessible degrees of freedom. At the same time, the new inputs are used to compensate for the dynamic coupling effects introduced by the coexistence of legacy and auxiliary actuation.
The approach leads to a process which we defined as  \emph{re-targeting} of such platform.

The remainder of this letter is organized as follows.  Sec.~\ref{sect:mot_exmp} introduces a motivating example.  Sec.~\ref{prelim}  presents some basic notions.
In Secs.~\ref{sect:probForm} and~\ref{sect:approach}, we formulate the problem and present the proposed solution, respectively.
The letter concludes with a continuation of the motivating example and corresponding simulations.

\section{Motivating Example}\label{sect:mot_exmp}
As a simple motivating example, consider a company operating a fleet of commercial quadrotors. These platforms, being underactuated, cannot achieve full 6D motion such as pure tilt without translation or pure translation without tilt, (see Fig.~\ref{fig:quad}). While one option would be to purchase a fully actuated aerial platform, such systems are scarce, expensive, and less practical. A more convenient solution is to customize an existing quadrotor, which preserves the certified hardware and software, along with the safety and usability features already in place. By adding two additional propellers, the platform gains two extra inputs, enabling 6D dexterity. However, the four original inputs of the commercial platform correspond to high-level commands (e.g., velocity and yaw rate) rather than direct propeller speeds, and modifying the internal controller is not feasible for legal and economic reasons. The resulting problem is therefore to design an outer-loop controller that leverages all six inputs while keeping the internal underactuated controller intact.

\section{Preliminaries}\label{prelim}
For a comprehensive introduction to feedback linearization, the interested reader is referred to \cite{1995-Isi}.

Consider a multi-variable nonlinear system
    \begin{equation}
   \left\{ \begin{split}
     \dot{\xv}&= \fv(\xv)+\Gm(\xv)\uv\\
    \yv &= \hv(\xv),
    \label{eq:sys}
    \end{split}\right.
    \end{equation}
where $\xv\in \mathbb{R}^n$ is the state, $\uv\in \mathbb{R}^p$ is the control input,  $\Gm(\xv)=\begin{bmatrix}
    \gv_1(\xv)  \cdots  \gv_{p}(\xv)
\end{bmatrix}\in \mathbb{R}^{n \times p}$, $\fv(\xv),\gv_1(\xv),\ldots,\gv_{p}(\xv)$ are smooth vector fields, and \mbox{$\hv(\xv)=\begin{bmatrix} h_1(\xv)  \cdots h_{p}(\xv)\end{bmatrix}^\top$} is a smooth function defined on an open set of $\mathbb{R}^n$.
The  system (\ref{eq:sys}) is said to have \emph{(vector) relative degree} $\rv = \{
    r_1, \ldots, r_{p}
\}$ at a point $\xv^\circ$ w.r.t. the input-output pair  $(\uv,\yv)$ if  
\begin{flalign}
&\text{\textrm{(i)}}    & L_{\gv_j}L^{k}_{\fv} h_i(\xv) &=0,&
\end{flalign}
for all $1\leq j \leq p$, for all $k< r_i-1$, for all $1\leq i \leq p$ and for all $\xv$ in a neighborhood of $\xv^\circ$, and\\
\textrm{(ii)}\;\;  the $p\times p$ matrix 
\begin{align}
        \Am(\xv)&:=\left [  \begin{smallmatrix}
            L_{\gv_1}L^{r_1-1}_{\fv}h_1(\xv) & \cdots&  L_{\gv_{p}}L^{r_1-1}_{\fv}h_1(\xv) \\ 
            L_{\gv_1}L^{r_2-1}_{\fv}h_2(\xv) & \cdots&  L_{\gv_{p}}L^{r_2-1}_{\fv}h_2(\xv) \\  
            \vdots & \ddots & \vdots \\
            L_{\gv_1}L^{r_{p}-1}_{\fv}h_{p}(\xv) & \cdots&  L_{\gv_{p}}L^{r_{p}-1}_{\fv}h_{p}(\xv) 
    \end{smallmatrix}\right  ]
    \label{eq:intbmatrix}\;,
\end{align}
is nonsingular at $\xv = \xv^\circ$.
 
 The integer $r_i$ denotes the number of times the $i$-th output $y_i(t)$ must be differentiated with respect to time, evaluated at $t = t^\circ$, until at least one component of the input vector $\uv(t^\circ)$ appears explicitly in the resulting expression.

At this point the generic expression of the output vector at the $\rv$-th derivative may be rewritten as an affine system of the form
\begin{equation}
\small
{\yv}^{(\rv)} :=\left[\begin{smallmatrix}
    y_1^{(r_1)} & \cdots & y_{p}^{(r_{p})}
\end{smallmatrix}\right]^\top =\bv(\xv)+\Am(\xv)\uv.
\label{eq:y_r_now}
\end{equation}
with 
\begin{equation}
 \small\bv(\xv):=\left[\begin{smallmatrix}
L_{\boldsymbol{f}}^{(r_1)}{h_1(\xv)} & \cdots & 
L_{\boldsymbol{f}}^{(r_{p})}{h_{p}(\xv)}
\end{smallmatrix}\right]^\top
\label{eq:b}
\end{equation}
and $\Am(\xv)$ as in~(\ref{eq:intbmatrix}).

Suppose the system \eqref{eq:sys} has some (vector) relative degree $\rv:=\{r_1,\ldots,r_p\}$ at $\xv^\circ$ and that the matrix $\Gm(\xv^\circ)$  has rank $p$ in a  neighborhood $\mathcal{U}$ of $\xv^\circ$. Suppose also that  \mbox{$r_1+r_2+\cdots+r_p=n$}, and choose the  control input to be 
\begin{equation}
\uv = \Am^{-1}(\xv)[-\bv(\xv)+\wv],
\label{eq:u2}
\end{equation}
where $\wv \in \mathbb{R}^{p}$ can be assigned freely.
Then, the output dynamics \eqref{eq:y_r_now}  take the form $\yv^{(\rv)} = \wv.$

\section{Problem Formulation}
\label{sect:probForm}

We consider a class of interconnected nonlinear systems composed of four main subsystems. The first is the \emph{plant} system, described by the state $\zv \in \mathbb{R}^s$:
\begin{equation}
\small
\left\{
    \begin{split}
\dot{\zv}&=\dv(\zv)+ \Gm_s(\zv)\cv_1 + \Gm_t(\zv)\cv_2\\
 \yv_1&=\left[
 h_1(\zv)\;\ldots\;
 h_{p_1}(\zv)
\right]^\top
\end{split}
\right. 
\label{eq:Sigma_p}\;,
\end{equation} where $\cv_1 \in \mathbb{R}^{p_1}$, and  $\cv_2 \in \mathbb{R}^{p_2}$. The functions $\dv$, $\Gm_s$, $\Gm_t$, and $h_i$ are smooth vector- or matrix-valued maps. The matrices $\Gm_s(\zv) \in \mathbb{R}^{s \times p_1}$ and $\Gm_t(\zv) \in \mathbb{R}^{s \times p_2}$ have full column rank.

The second subsystem is the \emph{primary actuation system}, with state $\xiv_{a_1} \in \mathbb{R}^{k_1}$:
\begin{equation}
\small
\left\{
   \begin{split}
\dot{\xiv}_{a_1}&=\varphiv_1(\xiv_{a_1})+ \Gm_{a_1}(\xiv_{a_1}) \vv_1 \\
\cv_1 &= \alphav_1(\xiv_{a_1}) + \betav_1(\xiv_{a_1}) \vv_1
\end{split}\right.\;,
\label{eq:Sigma_a_1}
\end{equation} 
where $\vv_1\in \mathbb{R}^{p_1}$,  and $\varphiv_1$, $\alphav_1$, $\Gm_{a_1}$,$\betav_1$ are smooth vector- or matrix-valued maps of appropriate dimensions.

The composite system formed by \eqref{eq:Sigma_p} and \eqref{eq:Sigma_a_1}, with output vector $\yv_1$ and $\cv_2=\bm{0}$, has a \emph{(vector)} relative degree \mbox{$\rhov_1:=\{r_1,\ldots,r_{p_1}\}$} at the point $(\zv,\xiv_{a_1})=(\zv^\circ,\xiv_{a_1}^\circ)$ and the sum $\rho_1 = r_1+r_2+\ldots +r_{p_1}$ equals  $s+k_1$.
Let $\bv_1(\zv, \xiv_{a_1})$ and $\Am(\zv, \xiv_{a_1})$ denote, respectively, the drift vector and the interaction matrix at this relative degree.  In particular, let us denote by $\mathcal{U}(\zv^\circ,\xiv^\circ_{a_1})$ the neighborhood around ($\zv^\circ,\xiv^\circ_{a_1}$) in which the matrix $\Am(\zv, \xiv_{a_1})$ is nonsingular.

Modern commercial platforms often include embedded controllers that come with pre-certified safety, stability, or performance guarantees. These internal controllers are not meant to be modified or re-tuned by end users. To formalize this setting, we introduce the following Defs.: 

\begin{defn}\label{def:1}
\emph{Internal controller certification} refers to the regulatory, safety, or performance guarantees associated with a pre-existing control module, assumed to be used as provided and without modification.
\end{defn}

\begin{defn}\label{def:2}
A control architecture is said to \emph{preserve the internal controller} if its structure and parameters remain unchanged, interaction occurs solely via certified input-output interfaces, and usage conforms to the platform’s certified operational scope.
\end{defn}

We now describe the internal controller subsystem, which is assumed to be a pre-certified component in the sense of Defs.~\ref{def:1}–\ref{def:2} and must be used as provided. In particular, it is defined as : 
   \begin{equation}
   \small
\left \{ \begin{split}\vv_1 
&= \Am(\zv,\xiv_{a_1})^{-1}\left[\tilde{\uv}_1-\bv_1(\zv,\xiv_{a_1}) \right]\\
\tilde{\uv}_1 &= {\yv^{(\rhov_1)}_{1,r}+ \sum^{\rhov_1-1}_{i=1}\Km_i(\yv^{(i)}_{1,r}-\yv^{(i)}_1)}
\end{split}\right.\,\,,
          \label{eq:Sigma_c}
          \end{equation}
{where the reference signals $\yv_{1,r}^{(i)}, i=1, \ldots, \rhov_1,$ are assumed to be externally generated and available for feedback, and $\Km_i>0$ are fixed, \textit{unknown}, positive-definite gain  matrices. The controller \eqref{eq:Sigma_c} is well-posed provided the pair $(\mathbf{z}, \xiv_{a_1})$ remains in the subset $\mathcal{U}(\zv^\circ, {\xiv}^\circ_{a_1})$ where $\Am(\zv, {\xiv}_{a_1})$ is nonsingular. }

{To remove the dependence on the internal gains, we define a reference assignment: 
\begin{equation}
\begin{split}
\yv^{(i)}_{1,r} &= \yv^{(i)}_1, \quad i=1,\ldots ,\rhov_1-1, \quad
\yv^{(\rhov_1)}_{1,r}=\uv_1,
\end{split}
\end{equation}
where $\uv_1$ is freely chosen. Under this choice, all error terms in the feedback law vanish except the highest-order one, yielding $\tilde{\uv}_1 = \uv_1$ and thus \mbox{$\small 
\vv_1 = \Am(\zv,\xiv_{a_1})^{-1}[\uv_1 - \bv_1(\zv,\xiv_{a_1})],
$}
which  cancels the influence of the internal controller and exposes $\uv_1$ for outer-loop design. Crucially, this reference assignment leaves the internal controller’s structure and parameters unchanged, interacts only via certified I/O channels, and respects its intended usage—thus preserving it in the sense of Def.~\ref{def:2}. 
}

The composite system defined by \eqref{eq:Sigma_p}, \eqref{eq:Sigma_a_1}, and  \eqref{eq:Sigma_c}, with $\cv_2=\bm{0}$,  represents a  commercial system equipped with its own internal controller. The  vector $\uv_1$ serves as  a  \emph{virtual} input vector, corresponding to the   $\rhov_1$-derivatives of the output vector $\yv_1$.

The fourth subsystem is the \emph{auxiliary actuation system}, with state $\xiv_{a_2} \in \mathbb{R}^{k_2}$:
\begin{equation}
\small
\left\{
\begin{aligned}
\dot{\xiv}_{a_2} &= \varphiv_2(\xiv_{a_2}) + \Gm_{a_2}(\xiv_{a_2}) \uv_2 \\
\cv_2 &= \alphav_2(\xiv_{a_2}) + \betav_2(\xiv_{a_2}) \uv_2
\end{aligned}
\right.\;,
\label{eq:Sigma_a}
\end{equation}
where $\uv_2 = [u_{p_1+1} \cdots u_p]^\top$, with $p:= p_1 + p_2$  and where $\varphiv_2$, $\alphav_2$, $\Gm_{a_2}$, and $\betav_2$ are smooth vector- or matrix-valued maps of appropriate dimensions.

Consider an \textit{auxiliary} output vector \begin{equation}
    \yv_2 = \begin{bmatrix}
        h_{p_1+1}(\zv)\;
        \cdots \;
        h_p(\zv)
\end{bmatrix}^\top=:\hv_2(\zv)\,\,,
\end{equation}
where $h_{p_1+1}(\zv),\ldots, h_{p}(\zv)$ are smooth functions, defined on an open set of $\mathbb{R}^{s}$.
To model the integration of the auxiliary actuation system with the original system, we substitute the output of \eqref{eq:Sigma_a} into \eqref{eq:Sigma_p} as follows:

\begin{equation}
\small
\left\{\begin{split}
\dot{\zv}&=\dv(\zv)+\Gm_{s}(\zv)\cv_1 + \Gm_{t}(\zv)\cv_2 \\
\dot{\xiv}_{a_1}&=\varphiv_1(\xiv_{a_1})+\Gm_{a_1}(\xiv_{a_1})
    \vv_1\\
    \dot{\xiv}_{a_2}&=\varphiv_{2}(\xiv_{a_2})+\Gm_{a_2}(\xiv_{a_2})
    \uv_2\\
    \cv_1&=\alphav_1(\xiv_{a_1})+\betav_1(\xiv_{a_1})\vv_1\\
        \cv_2&=\alphav_2(\xiv_{a_2})+\betav_2(\xiv_{a_2})\uv_2\\
        \vv_1 
&= \Am(\zv,\xiv_{a_1})^{-1}\bigl[\uv_1 - \bv_1(\zv,\xiv_{a_1})\bigr]
\end{split}\right.\,,
\label{eq:Sigma_ioext}
\end{equation}


For compactness, the system can be represented as

\begin{equation}
\small
\left\{
\begin{split}
     \dxv &= \fv(\xv)+\Gm_1(\xv)\uv_1+\Gm_2(\xv)\uv_2 \\
      \yv&=\hv(\xv)
\end{split}\;,
\right.
\label{eq:compact_system}
\end{equation}
with state $\xv=\begin{bmatrix}
    \xv_1^\top  \; \xv_2^\top 
\end{bmatrix}^\top \in \mathbb{R}^n $
 having set  \mbox{$
    \xv_1 = \begin{bmatrix}
    \zv^\top \; \xiv_{a_1}^\top 
\end{bmatrix}^\top$,} $
\xv_2 = \xiv_{a_2}$,$
\uv= \begin{bmatrix}
    \uv_1^\top  \; \uv_2^\top
\end{bmatrix}^\top$, and the overall output vector \mbox{$
\yv=[\yv_1^\top\;\yv_2^\top]^\top$}.

In the remainder of this letter, we write $h_i(\zv)$ instead of $h_i(\xv_1)$ for notational convenience, since $h_i$ depends only on $\zv \subset \xv_1$.

The drift vector is defined as $
\fv(\xv):=\fv_1(\xv_1)+\fv_2(\xv)$ with
        \mbox{$\fv_1(\xv_1)=\left[\begin{smallmatrix}
\dv(\zv)+\Gm_{s}(\zv)\alphav_1(\xiv_{a_1})-\Gm_{s}(\zv)\betav_1(\xiv_{a_1})\Am(\zv,\xiv_{a_1})^{-1}\bv_1(\zv,\xiv_{a_1})\\
        \varphiv_1(\xiv_{a_1}) - \Gm_{a_1}(\xiv_{a_1})\Am(\zv,\xiv_{a_1})^{-1}\bv_1(\zv,\xiv_{a_1})\\
       \bm{0}
\end{smallmatrix}\right]$}, and \mbox{$ \fv_2(\xv)=\left[\begin{smallmatrix} \Gm_{t}(\zv)\alphav_2(\xiv_{a_2})\\
\bm{0}\\
\varphiv_2(\xiv_{a_2})
\end{smallmatrix}\right]$}. The input matrices are$$\Gm_1(\xv)=\left[\begin{smallmatrix}  \Gm_{s}(\zv)\betav_1(\xiv_{a_1})\Am(\zv,\xiv_{a_1})^{-1} \\
\Gm_{a_1}(\xiv_{a_1})\Am(\zv,\xiv_{a_1})^{-1} \\
\bm{0}
\end{smallmatrix}\right] 
\Gm_2(\xv)=\left[\begin{smallmatrix}  \Gm_{t}(\zv)\betav_2(\xiv_{a_2}) \\
\bm{0}\\
\Gm_{a_2}(\xiv_{a_2}) \\
\end{smallmatrix}\right],$$
 of dimensions $n \times p_1$ and $n \times p_2$, respectively.

{The analysis assumes $(\zv^\circ, \xiv_{a_1}^\circ) \in \mathcal{U}(\zv^\circ, \xiv_{a_1}^\circ)$, a neighborhood where $\Am(\zv, \xiv_{a_1})  $ is nonsingular. Thus, $ \fv_1(\xv_1) $ and $ \Gm_1(\xv^\circ) $  are well-defined, with $  \Gm_1(\xv^\circ)$ having constant rank $p_1$ in this region.}

Throughout the remainder of the paper, we use \( \gv_j(\xv) \) to denote the $j$-th column of the matrix \mbox{\( \Gm(\xv) = [\Gm_1(\xv) \; \;\Gm_2(\xv)] \)}.

\begin{problem}\label{prob:main_problem}
Derive conditions on the  system \eqref{eq:compact_system} under which  there exists a  compensator  of the form \begin{equation}
\uv= \boldsymbol{\kappa}(\xv)+\etav(\xv)\wv\,\,, 
\label{eq:Sigma_o}
\end{equation}
 such that the composite system \eqref{eq:compact_system}, \eqref{eq:Sigma_o},  is input-state-output
linear and decoupled, i.e., represented by $p$ chains of integrators between $\wv$ and $\yv$ of the form
\begin{equation}
    \yv^{(\rhov)} = \wv\,\,,
    \label{eq:final_output_dynamics}
\end{equation} 
where $\wv \in \mathbb{R}^{p}$ can be assigned freely  and $\rhov$ denotes the (vector) relative degree.
\end{problem}
We refer to  Problem \ref{prob:main_problem} as the \textit{input}-\textit{output} extension problem, because the original commercial system is extended both in its input capability (through the \emph{auxiliary actuation system}) and output requirements (through the the \emph{auxiliary output vector}). What makes this problem nontrivial is: 1) the inability to access the \emph{primary actuation system} directly (no direct access to $\vv_1$) and the obligation to use the virtual input $\uv_1$ provided by the internal controller instead, and 2) the  fact that the internal controller operates assuming $\cv_2=\bm{0}$ and thus introduces a coupling between the primary and auxiliary outputs when the auxiliary actuation system is used ($\cv_2\neq\bm{0}$).

\section{Approach}
\label{sect:approach}
Having formalized the input-output extension problem in Section \ref{sect:probForm},  we will present the conditions on \eqref{eq:compact_system} under which a compensator can be synthesized.

\begin{thm}
\label{thm:1}
    If for the system \eqref{eq:compact_system} the following two conditions hold:
\begin{enumerate}
          \item     (Invariance of the (vector) Relative Degree of the Primary Output $\yv_1$) \label{cond_1}
          For all $1\leq i\leq p_1$  and for all $k < r_i-1$:
\begin{equation} 
L_{\gv_j} L_{\fv}^k h_i(\zv) = 0 \quad \forall j=1,\dots,p\,\,,
\end{equation}
i.e., the auxiliary actuation system does not modify the (vector) relative degree $\rhov_1$ of the preexisting  output vector $\yv_1$.

    \item (Nonsingularity of the Full Interaction Matrix) \label{cond_2}
 The interaction matrix $\Gammam(\xv) \in \mathbb{R}^{p \times p}$ defined as
   \begin{equation}
\scalebox{0.91}{$ 
\left [\begin{smallmatrix}
L_{\gv_1} L^{r_1 - 1}_{\fv} h_1(\zv) \ev_1^\top \Am^{-1}(\zv,\xiv_{a_1}) \ev_1 & \cdots & L_{\gv_p} L^{r_1 - 1}_{\fv} h_1(\zv) \\
\vdots & \ddots & \vdots \\
L_{\gv_1} L^{r_{p_1} - 1}_{\fv} h_{p_1}(\zv) \ev_{1}^\top \Am^{-1}(\zv,\xiv_{a_1}) \ev_{1} & \cdots & L_{\gv_p} L^{r_{p_1} - 1}_{\fv} h_{p_1}(\zv)\\
L_{\gv_1} L^{r_{p_1+1} - 1}_{\fv} h_{p_1+1}(\zv) \ev_{1}^\top \Am^{-1}(\zv,\xiv_{a_1}) \ev_{1} & \cdots & L_{\gv_p} L^{r_{p_1+1} - 1}_{\fv} h_{p_1+1}(\zv)\\
\vdots  & \ddots & \vdots\\
L_{\gv_1} L^{r_{p} - 1}_{\fv} h_{p}(\zv) \ev_{1}^\top \Am^{-1}(\zv,\xiv_{a_1}) \ev_{1} & \cdots & L_{\gv_p} L^{r_{p} - 1}_{\fv} h_{p}(\zv)\\
\end{smallmatrix} \right]
$},
\end{equation}
 is nonsingular for all $\xv$ in a neighborhood of the operating point \mbox{$\xv^\circ = (\zv^\circ,\xiv_{a_1}^\circ,\xiv_{a_2}^\circ)$}, with $\ev_j$ denoting a column vector with $1$ at position $j$ and zeros elsewhere.
\end{enumerate}
Then, a compensator guaranteeing the requirements specified in  Problem~1 can be designed.
\end{thm}
\begin{proof}
Under Cond.~\ref{cond_1}, and by evaluating the dynamics of the output $\yv_1$  at the (vector) relative degree $\rhov_1$, we obtain $p_1$ output equations:
\begin{equation} \resizebox{1\hsize}{!}{$ \begin{split} y_i^{(r_i)} &=L_{\fv_1}^{r_i} h_i(\zv)+\sum_{j=1}^{p_1} L_{\gv_j} L_{\fv_1}^{r_i - 1} h_i(\zv) \ev_j^\top (\Am(\zv,\xiv_{a_1})^{-1}[-\bv_1(\zv,\xiv_{a_1})+\uv_1])\\
&+L_{\fv_2}^{r_i} h_i(\zv)+ \sum_{j=1}^{p_1} L_{\gv_j} L_{\fv_2}^{r_i - 1} h_i(\zv) \ev_j^\top (\Am(\zv,\xiv_{a_1})^{-1}[-\bv_1(\zv,\xiv_{a_1})+\uv_1])\\&+ \sum_{j=p_1+1}^{p} L_{\gv_j} L_{\fv}^{r_i - 1} h_i(\zv) u_j =l_i(\xv)+\sum^{p}_{j=1}\gamma_{ij}(\xv)u_j, \end{split}$} \end{equation}

where the expressions of \(l_i(\xv)\) and \(\gamma_{ij}(\xv)\) follow directly from the above computation and are omitted for brevity. 
The first line reduces to $\ev_i^\top \uv_1 \ev_i$ since it corresponds to the action of the internal controller  while the second and third lines correspond to the contributions of $\fv_2$ and the remaining inputs, respectively.

Defining the vector-valued function \mbox{$\lv_1(\xv) := [{l}_1(\xv)\cdots{l}_{p_1}(\xv)]^\top$} and the decoupling matrix $\Gammam_1(\xv)$ accordingly, we can write compactly
\begin{equation}
\small
\yv_1^{(\rhov_1)} = \lv_1(\xv) + \Gammam_1(\xv) \uv.
\end{equation}
The remaining outputs at the (vector) relative degree $\rhov_2=\{r_{p_1+1},\ldots,r_{p} \}$ are:
\begin{equation}
\small
\yv_2^{(\rhov_2)} = \lv_2(\xv) + \Gammam_2(\xv) \uv,
\end{equation}
where $\lv_2(\xv)$ and $\Gammam_2(\xv)$ are defined analogously.
Concatenating both expressions yields:
\begin{equation}
\small
\yv^{(\rhov)}:=[(\yv_1^{(\rhov_1)})^\top\;(\yv_2^{(\rhov_2)})^\top]^\top = \lv(\xv) + \Gammam(\xv) \uv.
\end{equation}
with $\lv(\xv)=[\lv_1(\xv)^\top \; \lv_2(\xv)^\top]^\top$
Then, under {Condition \ref{cond_2}} , the feedback law
\begin{equation}
\small
\uv = \Gammam(\xv)^{-1} \left[ -\lv(\xv) + \wv \right],
\label{eq:ctrl_w_unk}
\end{equation}
 ensures that the problem is solvable. 
\end{proof}
\begin{rem}
Condition~\ref{cond_1} preserves the vector relative degree $\rhov_1$ of the primary output $\yv_1$, ensuring the original input–output mapping and thus the validity of any certified internal controller without re-design. It can often be enforced via dynamic extension, as in the later example. In commercial platforms, internal control typically acts at a fixed output derivative (e.g., acceleration or snap), and the architecture prevents actuation from affecting lower levels (e.g., position).
\end{rem}

\begin{rem}
    The structure of \eqref{eq:Sigma_c} and \eqref{eq:Sigma_a} considers also the case of  static actuation systems $k_1=0$ and/or $k_2=0$.
\end{rem}

\begin{rem}
Condition~\ref{cond_2} is a function of the primary actuation state \(\xiv_{a_1}\); in practice, its evaluation may require additional sensing or state estimation. 
\end{rem}

\begin{rem}
    The  control law \eqref{eq:ctrl_w_unk} is guaranteed to preserve the internal controller structure in the sense of Def.~\ref{def:2}.
\end{rem}

\subsection{Trajectory Tracking}

    Consider a  desired  output trajectory \mbox{$\yv_d:(0,\infty)\to \mathbb{R}^p$} and its derivatives. Define the error $\ev :=\yv_d-\yv.$ If Problem 1 is solvable, then, in order to exponentially stabilize the origin of the error dynamics $\ev$, we can select the  input vector $\wv$ as
\begin{equation}
\small
    \wv = \yv_{d}^{(\rhov)} + 
    \sum\nolimits_{i=0}^{\rhov-1} \Lm_i \left({\yv}^{(i)}_d - {\yv}^{(i)}\right),
    \label{eq:control2}
\end{equation} where  the $\Lm_i$ matrices are chosen arbitrarily, subject only to the constraint that substituting~\eqref{eq:control2} into~\eqref{eq:ctrl_w_unk} results in a linear output error dynamics that is exponentially stable.

\section{Application to the Motivating Example}\label{sect:app_mot_exmp}
To demonstrate the practical utility of Theorem~\ref{thm:1}, we apply it to the quadrotor platform introduced in Section~\ref{sect:mot_exmp} and displayed in Fig.~\ref{fig:quad}. 
We define an inertial frame \( \mathcal{F}_W = \{O_W, \xv_W, \yv_W, \zv_W\} \) and a body-fixed frame \mbox{$ \mathcal{F}_B = \{O_B, \xv_B, \yv_B, \zv_B\}$}, rigidly attached to the platform. The origin $O_B$, located at the CoM and geometric center of the six propellers, has position \( \pv_R \in \mathbb{R}^3 \) and orientation \( \Rm \in \SO \). The angular velocity of the platform w.r.t. $\mathcal{F}_W$, expressed in $\mathcal{F}_B$, is \( \Omegav \in \mathbb{R}^3 \), and the rotation matrix satisfies \( \dot{\Rm} = \Rm \Omegav^\times \), where \( \av^\times \in \mathfrak{so}(3) \) denotes the skew-symmetric matrix of \( \av \in \mathbb{R}^3 \).
Assuming that no task for {the platform requires to fly with the sagittal $\xv_B$ axis pointing down or up like $\pm \zv_W$}, we parametrize the rotation matrix as $\Rm = \Rm(\Phim)$, where \mbox{$\Phim = [\phi\;\theta\;\psi]^\top$} are the roll-pitch-yaw (RPY) angles.
The body angular velocity is related to the RPY rates by:
\begin{equation}
\Omegav = \Tm(\Phim)\dot{\Phim}, \quad
\Tm(\Phim) = \left[\begin{smallmatrix}
1 & 0 & -\sin(\theta) \\
0 & \cos(\phi) & \cos(\theta)\sin(\phi) \\
0 & -\sin(\phi) & \cos(\theta)\cos(\phi)
\end{smallmatrix}\right].
\end{equation}

The equations of motion for the platform are given compactly by:
\begin{equation}
\small
\left\{
\begin{aligned}
\dot{\zv} &= 
\left[\begin{smallmatrix}
\vv \\
-\!g\ev_3 + \frac{1}{m} \Rm(f_1\ev_1 + f_2\ev_2 + f_3\ev_3) \\
\Tm^{-1}(\Phim)\Omegav \\
-\Jm^{-1} \Omegav^\times \Jm \Omegav + \Jm^{-1} \tauv
\end{smallmatrix}\right], \quad
\yv_1 = 
\left[\begin{smallmatrix}
\vv^\top & \psi
\end{smallmatrix}\right]^\top
\end{aligned},
\right.
\label{eq:sigma_p_quad}
\end{equation}
where   $\zv = \left[\begin{smallmatrix} \pv_R^\top & \vv^\top & \Phim^\top & \Omegav^\top \end{smallmatrix}\right]^\top$, is the state,  $g$ is the gravitational constant, \( \ev_3 = [0\;0\;1]^\top \), \( m \) is the platform’s total mass, and \( \Jm \in \mathbb{R}^{3\times 3}\) is the inertia matrix. The scalar $f_3$ and the vector $\tauv \in \mathbb{R}^3$ denote the thrust force and the  torque inputs, respectively. 
The scalars $f_1$ and $f_2$ correspond to the two lateral forces produced by the additional propellers.  

The \emph{primary actuation system} has  state \mbox{\(\xiv_{a_1} = [f_3~\dot{f}_3]^\top\)} described by  
\begin{equation}
\small
\left\{
    \begin{split}
       \dot{\xi}_{a_1,1} &= \xi_{a_1,2} \\ 
       \left[\begin{smallmatrix}
           \dot{\xi}_{a_1,2}\;
           \tauv^\top
       \end{smallmatrix}\right]^\top & =\vv_1, \quad
       \cv_1 = \left[\begin{smallmatrix}
           \xi_{a_1,1}\;
           \tauv^\top
        \end{smallmatrix}\right]^\top
    \end{split}\;.
\right.
\label{eq:sigma_a_1_quad}
\end{equation}
Let the velocity and yaw angle tracking errors be defined as \(\ev_v := \dot{\pv}_{R,r} - \dot{\pv}_R\), and \(\ev_\psi := \psi_r - \psi\), respectively.
Consider the composite system \eqref{eq:sigma_p_quad},\eqref{eq:sigma_a_1_quad} and denote with \(\Am(\xv_1)\)   the interaction  matrix and with \(\bv(\xv_1)\)  the drift term
at the relative degree \({\rhov}_1 = \{3, 3, 3, 2\}\) where $\xv_1 =[\zv^\top \; \xiv_{a_1}^\top]^\top$.  We omit, for brevity, the explicit expressions of \(\bv(\xv_1)\) and \(\Am(\xv_1)\), which can be found in~\cite{5164788}.

The internal controller is given by the feedback law:
\begin{equation}
   \small \left\{
\begin{split}
\vv_1&= \Am(\xv_1)^{-1}[-\bv(\xv_1) + \tilde{\uv}_1]\\
\tilde{\uv}_1&=\left [\begin{smallmatrix}\pv_{R,r}^{(4)}\\ \; \ddot{\psi}_r\end{smallmatrix} \right]+\left [\begin{smallmatrix}
    \Km_1 \ev_v+\Km_2 \dot{\ev}_v+\Km_3 \ddot{\ev}_v\\
k_{p\psi}e_\psi+k_{d\psi}\dot{e}_\psi \end{smallmatrix} \right]
\end{split}
\right.
\label{eq:sigma_c_in_quad}\,,
\end{equation}
where \mbox{$k_{p\psi}>0,$} $k_{d\psi}>0$, $\Km_1,\Km_2,\Km_3>0$ are not-tunable gains.
The controller is well defined as long as  the working point $(f_3^\circ,\phi^\circ)$ is such that $(f_3^\circ)^2 \cos(\phi^\circ) \neq 0$, which is standard and ensures non-degenerate thrust control—especially for commercial systems.

{ The virtual inputs from the internal controller are assigned as follows:}
\begin{equation}
\small
\left\{
    \begin{split}
[\dpv_{R,r}^\top\;\ddpv_{R,r}^\top\;\dddot{\pv}_{R,r}^\top]^\top&=[\dpv_R^\top\;\ddpv_R^\top\;\dddot{\pv}_R^\top]^\top,  \;    
[\psi_{r}\;\dot{\psi}_r]^\top=[\psi\;\dot{\psi}]^\top\\
    [(\pv_{R,r}^{(4)})^\top \ddot{\psi}_r]^\top &= {\uv}_1
    \end{split} 
    \right.
    \label{eq:assignment_gain}\,,
\end{equation}
where $\uv_1$ is a freely chosen input.

The composite system \eqref{eq:sigma_p_quad}, \eqref{eq:sigma_a_1_quad},
\eqref{eq:sigma_c_in_quad} with \mbox{$f_1=f_2=0$}  represents the commercial platform with its own controller exposing as virtual commands the ones corresponding to $\uv_1$. It is important to highlight that the internal controller is 1) given  (fixed) and 2) does not use nor consider the presence of the variables $f_1$ and $f_2$. The controller of the re-targeted platform will instead make use of those inputs together with the virtual commands $\uv_1$.

We define the \emph{auxiliary actuation system} with state  \mbox{$\xiv_{a_2} =[f_1 \; f_2 \; \dot{f}_1 \; \dot{f}_2]^\top$} and given by:
\begin{equation}
\small
\left\{ \begin{split}
\dot{\xiv}_{a_2}&=\left[\begin{smallmatrix}
    0 & 0 & 1 & 0\\
    0 & 0 & 0 & 1 \\
    0 & 0 & 0 & 0 \\
    0 & 0 & 0 & 0
\end{smallmatrix}\right]{\xiv}_{a_2}+\left[\begin{smallmatrix}
    0 & 0\\
     0 & 0\\
      1 & 0\\
       0 & 1\\
\end{smallmatrix}\right]\uv_2
\\
     \cv_2 &= \left[\begin{smallmatrix}
         1 & 0 & 0 & 0\\
         0 & 1 & 0 & 0
\end{smallmatrix}\right]
     \xiv_{a_2}
\end{split}\right.\;,
\label{eq:sigma_a_quad}
\end{equation}
where  $\uv_2$ are the \emph{additional} inputs. 

\begin{rem}
One may be tempted to choose $f_1$ and $f_2$ directly as inputs instead of their second derivatives. 
However, it can be proven that, in that case, the system would not satisfy the conditions of Theorem~\ref{thm:1}.
\end{rem}

By introducing these additional inputs through such actuation systems, we are able to control more variables beyond those accessible via the internal controller alone. Hence, we can consider an auxiliary output constituted by the roll and pitch angle respectively i.e.,  \( \yv_2 = [\phi\; \;\theta]^\top \). 

Let the overall state be given by \(\xv = [\xv_1^\top\; \xv_2^\top]^\top\), where \mbox{\(\xv_2 = \xiv_{a_2}\)}.
It can be shown that the composite system \eqref{eq:sigma_p_quad},\eqref{eq:sigma_a_1_quad},\eqref{eq:sigma_c_in_quad},\eqref{eq:assignment_gain},\eqref{eq:sigma_a_quad} with output vector $\yv =[\yv_1^\top\;\yv_2^\top]^\top$, satisfies the conditions of Theorem~\ref{thm:1}, with the (vector) relative degree \(\rhov = \{\rhov_1, \rhov_2\}\), where \mbox{\(\rhov_2 = \{2,2\}\)}.
Accordingly, a feedback-linearizing controller can be constructed following the procedure outlined in the proof of Theorem~\ref{thm:1}.

\begin{figure*}[t]
    \centering
    \begin{minipage}[t]{0.48\textwidth}
        \centering
        \includegraphics[width=\textwidth]{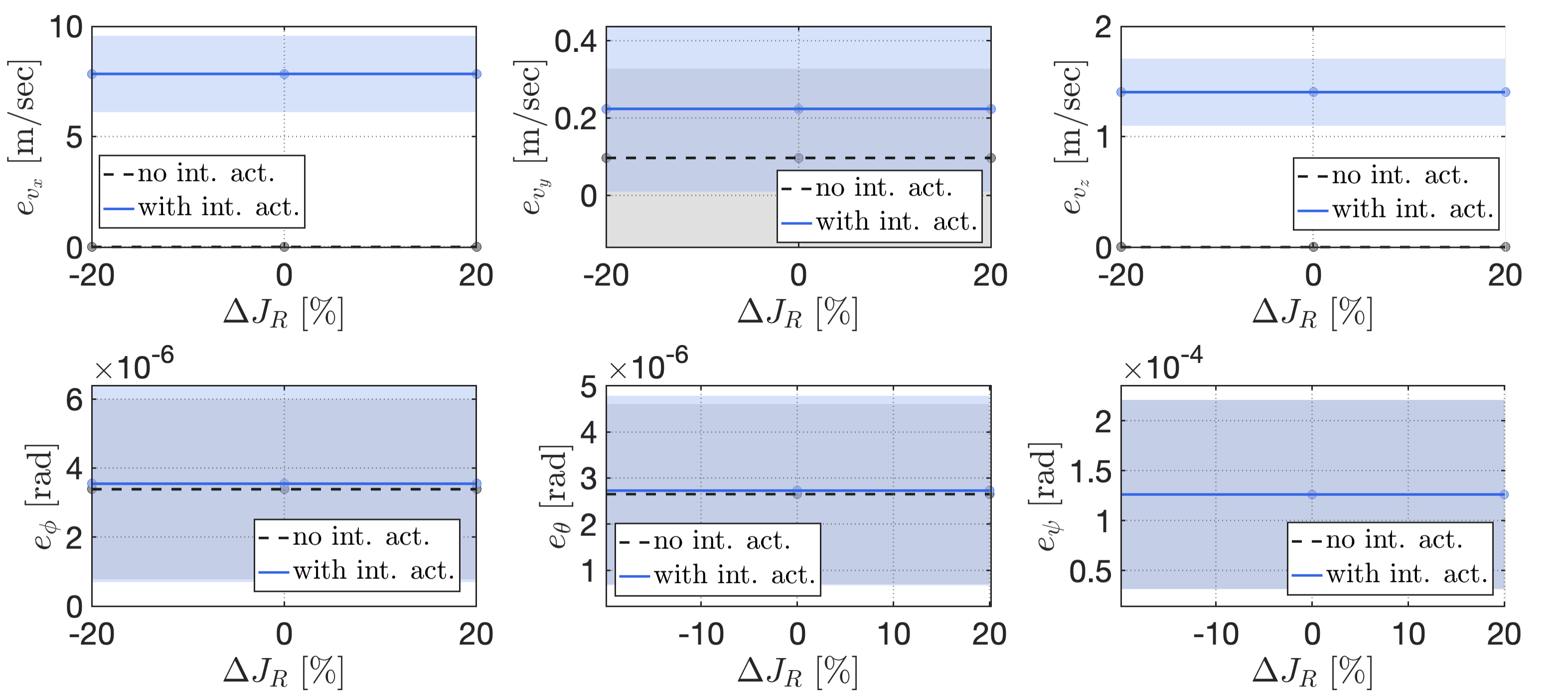}
        \captionof{figure}{Inertia uncertainty.}
        \label{fig:sim1a}
    \end{minipage}
    \hfill
    \begin{minipage}[t]{0.48\textwidth}
        \centering
        \includegraphics[width=\textwidth]{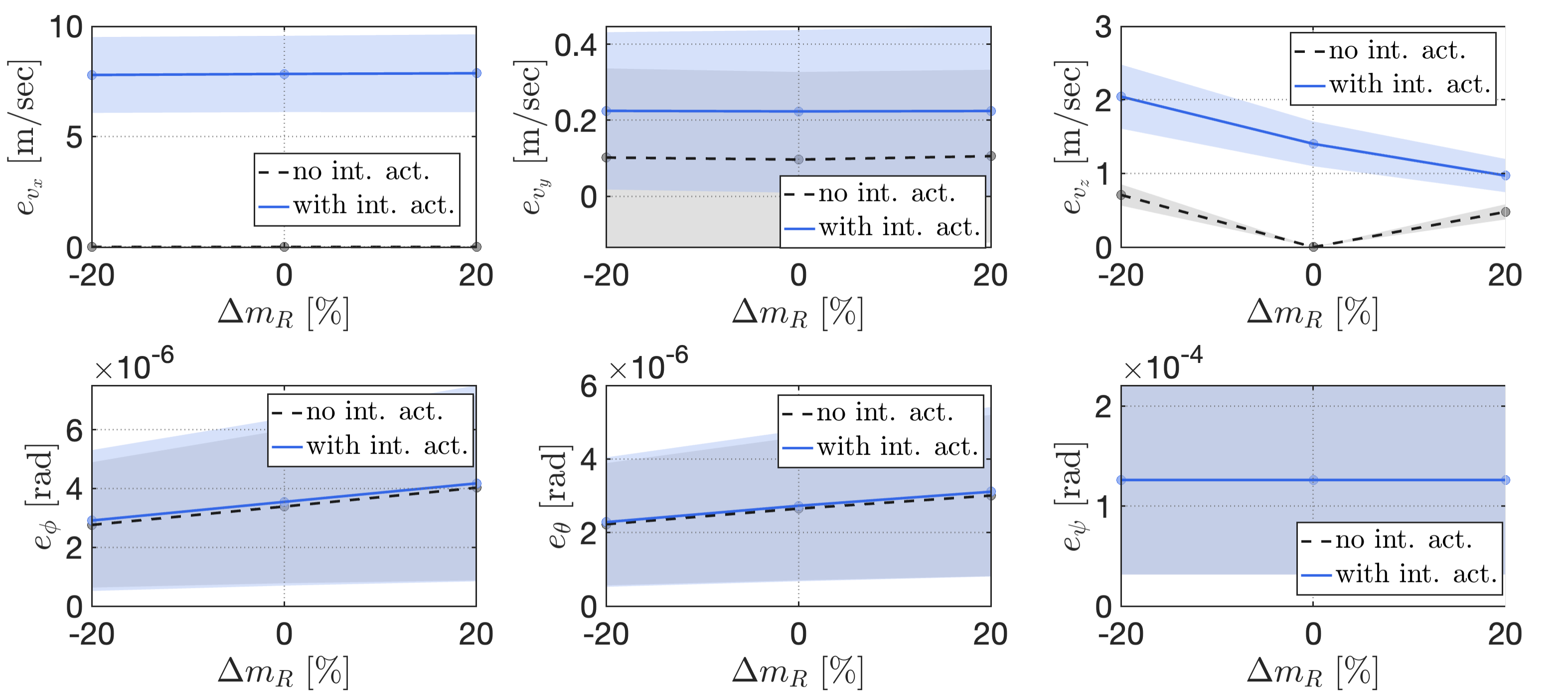}
        \captionof{figure}{Mass uncertainty.}
        \label{fig:sim1b}
    \end{minipage}
\caption{{\textbf{Scenario 1.} Robustness analysis under uncertainty, actuation noise, and reference perturbations. Solid lines represent the mean tracking error with integral action (blue), while dashed lines indicate results without integral action (black). Shaded regions correspond to $\pm$ one standard deviation. Subplots show errors in linear velocities $\ev_{v}$ and angular orientations $e_\phi$, $e_\theta$, and $e_\psi$.}}
    \label{fig:sim1}
\end{figure*}

\subsection{Numerical Simulations} 
{
The proposed control scheme is evaluated through simulations on two representative  tasks. To approximate real-world conditions, low-pass filtered Gaussian noise is added to the actuation inputs. Robustness is further tested under parametric uncertainties and internal reference perturbations.}

\subsubsection{Simulation Setup}
{
Noise is modeled as a low-pass filtered Gaussian process \( \nv \in \mathbb{R}^6 \), defined by $
    \dot{\nv} = -k \nv + \boldsymbol{\mu}$,  \mbox{$\boldsymbol{\mu} \in \mathcal{N}(\mathbf{0}, q^2 \mathbf{I}_6)$},
where $q = 0.4$ and $k = 0.1$.
To simulate implementation-level imperfections such as discretization, estimation errors, and delay, we perturb the internal reference assignment in the controller. Specifically, bounded disturbances are injected into the internal virtual output references:
\begin{equation}
\small
\yv^{(i)}_{1,r} = \yv^{(i)}_1 + \Delta_i, \quad i = 0, \ldots, \rhov_1 - 1,\;\;
\yv^{(\rhov_1)}_{1,r} = \uv_1 + \Delta_{\rhov_1},
\end{equation}
where \( \Delta_{i} \sim \mathcal{U}(-\varepsilon_i, \varepsilon_i) \) with \( \varepsilon_i \ll 1 \). These perturbations are applied consistently in all simulation scenarios.
Parametric robustness is evaluated by introducing mismatches between the true system parameters and those assumed in the controller. The true mass $m$ and inertia $\Jm$ are expressed as
$m = (1 + \Delta m)\, m_n, \quad \Jm = (\mathbf{I} + \Delta \Jm)\, \Jm_n,$
with  $ \Delta m \in \mathbb{R} $, symmetric \( \Delta \Jm \in \mathbb{R}^{3 \times 3} \), and bounds \mbox{\( |\Delta m| < 1 \)}, \mbox{\( \|\Delta \Jm\| < 1 \)} to maintain physical feasibility.
The values of  $\varepsilon_i$ were chosen as  $0.001$, for all $i$.
The quadrotor parameters are $ m = \SI{2.25}{\kilogram}$, $ \Jm = \SI{0.0207}{\kilogram\meter\squared} \, \mathbf{I}_3 $. The initial conditions are
$\pv_R(0) = \left[\begin{smallmatrix} 0 \; 0 \; 10 \end{smallmatrix}\right]^\top~\si{\meter}$, \mbox{$
\dot{\pv}_R(0) = \left[\begin{smallmatrix} 0 \; 1 \;0 \end{smallmatrix}\right]^\top~\si{\meter\per\second}$}, \mbox{$
\Omegav(0) = \left[\begin{smallmatrix} 0 \; 0 \; 0 \end{smallmatrix}\right]^\top~\si{\radian\per\second}$}, and $\Rm(0) = \mathbf{I}$.
Controller gains are selected as $
\Lm_1 = \left[\begin{smallmatrix} 30 \mathbf{I}_3 & \bm{0} \\ \bm{0} & 32 \mathbf{I}_3 \end{smallmatrix}\right]$, $
\Lm_2 = \left[\begin{smallmatrix} 20 \mathbf{I}_3 & \bm{0} \\ \bm{0} & 15\mathbf{I}_3 \end{smallmatrix}\right], $ and \mbox{$
\Lm_3 = \left[\begin{smallmatrix} 7.5 \mathbf{I}_3 & \bm{0} \\ \bm{0} & \bm{0} \end{smallmatrix}\right]$}.
Tracking performance is evaluated using the absolute tracking error \( e_i(t) = |y_i(t) - y_{d,i}(t)| \), its time average \( \bar{e}_i = \frac{1}{T} \sum_{t=1}^T e_i(t) \), and its standard deviation \mbox{$ \sigma_{\bar{e}_i} = \sqrt{ \frac{1}{T} \sum_{t=1}^T (e_i(t) - \bar{e}_i)^2 }$}. For angular signals, differences are wrapped using $
\mathrm{wrap}(\theta) = \mathrm{atan2}(\sin(\theta), \cos(\theta)).$
}

\subsubsection{Scenario 1: Lissajous-Type Path and Robustness}
{
This task consists of a spatially oscillatory trajectory in the horizontal plane, resembling a Lissajous curve, while maintaining constant altitude. It is defined as \mbox{$
\dot{\pv}_{R,d}(t) = \left[\begin{smallmatrix}-A\omega \sin(\omega t)\; A\omega \cos(2\omega t)\; 0\end{smallmatrix}\right]^\top$}, $
\Phim_d(t) =\mathbf{0}_{3},$
with \mbox{\( A = \SI{3}{\meter} \)}, \( \omega =  \SI{0.25}{\radian\per\second} \).
Robustness is assessed under combinations of parameter mismatch, internal reference perturbations, and actuation noise as defined in the setup. Figs~\ref{fig:sim1a}–\ref{fig:sim1b} report the results. A 10\% mass mismatch induces a steady-state error in the $\zv_W$-direction, which may be mitigated by incorporating integral action or  mass estimation. Inertia mismatch, by contrast, without integral action,  has negligible impact. 
}
\subsubsection{Scenario 2: Straight-Line Motion with Physical Interaction}
\begin{figure}[t]
    \centering\includegraphics[width=1.034\linewidth]{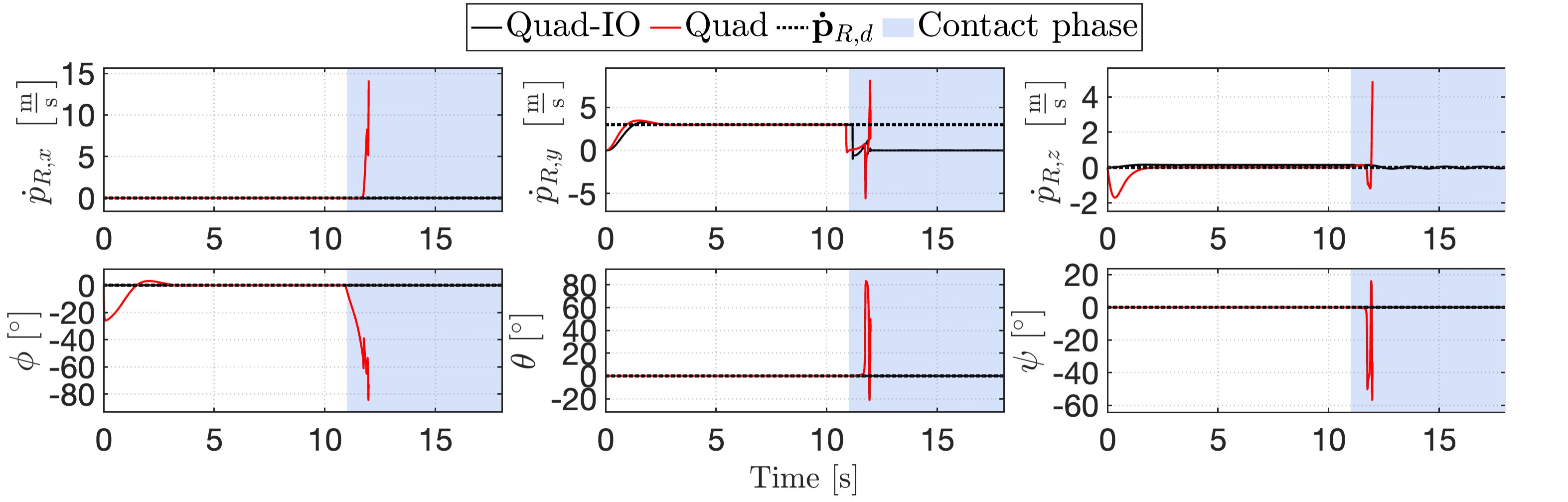}
      \caption{ {\textbf{Scenario 2.} The light blue region marks the contact phase. The quad-IO platform maintains stable tracking, while the standard quadrotor diverges after $11s$, highlighting an improvement in performances of the proposed approach. The roll and pitch evolution for the quadrotor are dictated by the flatness.}}
    \label{fig:sim2}
\end{figure}

{The second task involves a typical physical interaction with a wall. The platform is equipped with a lightweight, rigid end-effector (mass $m_E = 50,\text{g}$) mounted at the vehicle’s center of mass at position $^{B}\pv_E=[0\;0.5\;0]^\top$ to minimize its effect on the moment of inertia. The end-effector must exert a desired force on the wall while maintaining stable contact.  We modeled the contact using a nonlinear visco-elastic force model with stiffness $k_c$, damping $b_c$, and regularized friction. The resulting contact forces and torques, based on penetration depth and relative velocity, were integrated into the quadrotor dynamics.  The reference trajectory is a straight-line translation along the $\yv_W$ axis without rotation, defined by
\mbox{$\dot{\mathbf{p}}_{R,d}(t) = \left[0\;v_y\;0\right]^\top$, and $\Phim_d(t) = \mathbf{0}_{3},$}
where \mbox{$v_y=\SI{3}{\meter\per\second}$}. We use the same standard quadrotor platform with the mounted end-effector and evaluate its performance under two conditions: with the additional control module activated (referred to as \text{Quad-IO}) and without it (\text{Quad}). Both scenarios share identical initial conditions and parameters. Simulation results, shown in Fig.~\ref{fig:sim2}, indicate that without the additional module, the quadrotor is unable to successfully perform the interaction task, whereas enabling the module allows the platform to complete the task as intended.}

\section{Conclusion and Future Works}

This letter presents  a method to enhance the task-space capabilities of underactuated commercial platforms by integrating auxiliary actuation without altering their certified internal controllers. A feedback-linearizing outer-loop controller enables simultaneous tracking of both primary (e.g., velocity/yaw) and auxiliary (e.g., roll/pitch) objectives within a unified framework. Key features include:
\begin{enumerate}
\item Control Reinterpretation: Existing high-level commands are repurposed for unactuated degrees of freedom, while new inputs handle coupling.
\item Practical Viability: The method respects existing hardware/software constraints, preserving certification and safety, as shown in the quadrotor case.
\item Exponential Stability: Guaranteed under exact model knowledge.
\end{enumerate}
{
A current limitation is the requirement for knowledge of the internal controller, which must be provided or identified. Future work will investigate data-driven methods, including reinforcement learning, to relax this assumption and will validate the method experimentally on physical platforms.
}

\bibliographystyle{IEEEtran}
\bibliography{Settings/Bib/bibAlias,Settings/Bib/bibAlias_short,Settings/Bib/custom,Settings/Bib/bibAF,Settings/Bib/bibMain,Settings/Bib/bibNew}

\end{document}